\documentclass[aps,prm,reprint]{revtex4-1}
\usepackage{mathtools}
\usepackage{bm}
\usepackage{chemformula} 
\usepackage{charter}

\begin{document}

\title{Modeling magnetic evolution and exchange hardening in disordered magnets: 
The example of Mn$_{1-x}$Fe$_x$Ru$_2$Sn Heusler alloys}

\author{Elizabeth Decolvenaere}
\email{liz.decolvenaere@gmail.com}
\affiliation{Department of Chemical Engineering, University of California, Santa Barbara, California 93106, USA}
\affiliation{Materials Research Laboratory, University of California, Santa Barbara, California 93106, USA}

\author{Emily Levin}
\affiliation{Materials Department, University of California, Santa Barbara, California 93106, USA}
\affiliation{Materials Research Laboratory, University of California, Santa Barbara, California 93106, USA}

\author{Ram Seshadri}
\affiliation{Materials Department, University of California, Santa Barbara, California 93106, USA}
\affiliation{Materials Research Laboratory, University of California, Santa Barbara, California 93106, USA}

\author{Anton Van der Ven}
\email{avdv@ucsb.edu}
\affiliation{Materials Department, University of California, Santa Barbara, California 93106, USA}
\affiliation{Materials Research Laboratory, University of California, Santa Barbara, California 93106, USA}

\begin{abstract}
We demonstrate how exchange hardening can arise in a chemically-disordered solid solution from a 
first-principles statistical mechanics approach. A general mixed-basis chemical and magnetic cluster expansion 
has been developed, and applied to the Mn$_{1-x}$Fe$_x$Ru$_2$Sn Heusler alloy system; single-phase solid 
solutions between antiferromagnetic \ch{MnRu2Sn} and ferromagnetic \ch{FeRu2Sn} with disorder on the 
Mn/Fe sublattice that exhibit unexpected exchange hardening. Monte Carlo simulations applied to the cluster 
expansion are able to reproduce the experimentally measured magnetic transition temperatures and the bulk 
magnetization as a function of composition. The magnetic ordering around a site is shown to be dependent not 
only on bulk composition, but also on the identity of the site and the local composition around that site. The 
simulations predict that local antiferromagnetic orderings form inside a bulk ferromagnetic region at intermediate 
compositions that drives the exchange hardening. Furthermore, the antiferromagnetic regions disorder at a 
lower temperature than the ferromagnetic regions, providing an atomistic explanation for the 
experimentally-observed decrease in exchange hardening with increasing temperature. 
These effects occur on a length scale too small to be resolved with previously-used characterization 
techniques.
\end{abstract}

\pacs{}

\maketitle

\section{Introduction}

Magnetic exchange hardening, a phenomenon that increases a material's resistance to demagnetization, typically occurs in two-phase mixtures consisting of a ferromagnetic (FM) phase and an antiferromagnetic (AFM) 
phase.\cite{Skomski1993,Callsen2013,Pebley2017}
It is well established that magnetic hardening can emerge from magnetic interactions across the interfaces separating the FM and the AFM phases, which becomes especially pronounced in nano-structured two-phase mixtures. 
Less understood is the anomalous exchange hardening observed in the chemically disordered, single phase Heusler alloys Mn$_{1-x}$Fe$_x$Ru$_2$Sn and Mn$_{1-x}$Fe$_x$Ru$_2$Ge.\cite{Ishida1995,Douglas2016}
Full-Heusler alloys (which we refer to here simply as ``Heuslers'') consist of four interpentrating FCC-type sublattices, adopting a L2$_1$ chemical ordering.
In both Heuslers, the magnetic moments of the Mn-rich alloys prefer AFM ordering adopting an L1$_{1}$ ordering over the Mn sublattice, 
consisting of alternating FM planes along the [111] direction. The Fe-rich compositions display FM ordering of the magnetic moments.
Anomalous exchange hardening is observed at intermediate compositions, where the Fe/Mn sublattice of the Heusler 
alloys is chemically disordered. Prior experimental \cite{Ishida1995,Douglas2016} and 
\textit{ab-initio}\cite{Decolvenaere2017a} studies on these and similar systems\cite{Dutta2018} proposed hypotheses of short-range magnetic interactions to explain the observed single-phase Heusler exchange hardening, but were not able to directly demonstrate any local ordering phenomena.

Here, we show how exchange hardening can arise in a single chemically-disordered solid solution.
We develop a first-principles statistical mechanics model \cite{Sanchez1984, DeFontaine1994} that fully couples 
the chemical and magnetic degrees of freedom in the Mn$_{1-x}$Fe$_x$Ru$_2$Sn Heusler alloy and use it to predict 
magnetic short- and long-range order under experimentally-realistic 
conditions.\cite{VanderVen2008,VanderVen2010,Puchala2013}
We find that magnetic hardening in the disordered Heusler originates from \textit{local} AFM orderings, primarily localized on Mn sites, interacting with a bulk FM ordering.
The magnetic ordering around each site is very sensitive to small fluctuations in local composition.
As a result, quenched-in chemical disorder between Mn and Fe leads to a continuum of atomic scale AFM-in-FM magnetic orderings over a broad composition interval. 
This unusual behavior is akin to a Griffiths \cite{Griffiths1969, Burgy2001, Vojta2010, Wang2017} phase, 
and is responsible for the anomalous exchange hardening seen in this single-phase material.

\section{Theory and Methods}

Our approach uses a chemomagnetic cluster expansion Hamiltonian\cite{Sanchez1984, DeFontaine1994} that includes fully-coupled chemical and magnetic degrees of freedom. 
Prior work\cite{Douglas2016, Decolvenaere2017a} has shown that significant chemical disorder is present only on the (Mn,Fe) FCC sub-lattice, and that the moments on the Ru and Sn sites do not contribute meaningfully to the magnetic ordering.
The relevant chemical and magnetic configurational degrees of freedom can therefore be restricted to the (Mn,Fe) FCC sub-lattice, while the Ru and Sn sub-lattices are present only during the electronic structure calculations used to generate training data to parameterize the chemomagnetic cluster expansion.

The chemical and magnetic degrees of freedom in (Mn,Fe)Ru$_2$Sn can be described by assigning a pair of state variables to each site in the crystal \cite{Lavrentiev2010}: the site chemistry $x_i$, which is +1 when site $i$ is occupied by Fe and $-$1 when occupied by Mn, and the site moment $m_i$, which can also be $\pm 1$.
The state at each site $i$ is fully specified by the occupation vector $\bm{\sigma}_i = [x_i, m_i]$, while the state of the chemical and magnetic configuration of the crystal is specified by a microstate vector, $\overline{\sigma} = [\bm{\sigma}_0, \bm{\sigma}_1, \dots, \bm{\sigma}_n]$.
Any scalar property of configuration $\overline{\sigma}$ can be expanded in a basis of cluster functions $\phi^{(\gamma)}_\delta(\overline{\sigma})$ defined as\cite{Sanchez1984, DeFontaine1994} 
\begin{equation}
    \phi_\delta^{(\gamma)} (\overline{\sigma}) = \prod_{\substack{i \in \delta \\ (p,q) \in \gamma}}
    x_i^p m_i^q
    \label{phid}
\end{equation}
where $i$ denotes a site belonging to a cluster $\delta$ of sites in the (Mn,Fe) sublattice, such as a point, pair, or triplet of sites. 
The $p$ and $q$ are elements of a tuple of exponents determining which degrees of freedom participate in site $i$'s contribution to the cluster function and can be each 0 or 1; we will refer to the set $\gamma$ as the colors of the sites of the cluster.

    Many of the $\phi_\delta^{(\gamma)}(\overline{\sigma})$ are related to each other by symmetry operations.
All clusters $\delta$ that map onto a prototype cluster $\alpha$ by a symmetry operation of the crystal belong to the orbit $\Omega_\alpha$.
For each prototype cluster $\alpha$, there are several ways to color the sites of the cluster via $\gamma$.
All colourations $\gamma$ of the cluster $\alpha$ that are equivalent to $\beta$ under symmetry operations of the crystal belong to the orbit $\Omega_{\alpha}^{\beta}$.
Multiple unique prototype colourations $\beta$ exist for each $\alpha$.
A pair cluster, for example, has five unique cluster prototypes: $x_i x_j, m_i m_j, x_i m_i x_j, x_i m_i m_j$ and $x_i m_i x_j m_j$.

The energy (or any scalar property) of a microstate $\overline{\sigma}$ can be expressed as\cite{Sanchez1984}:
\begin{equation}
    E(\overline{\sigma}) = \sum_{(\alpha,\beta)} V_{(\alpha, \beta)} \sum_{(\delta, \gamma) \in \Omega_{\alpha}^{\beta}} \phi_\delta^{(\gamma)},\label{hamil}
\end{equation}
where the $V_{(\alpha, \beta)}$ are the expansion coefficients, referred to as effective cluster interactions.

    Hamiltonians accounting for magnetic degrees of freedom must not only satisfy crystal symmetries, but also time-reversal symmetry.\cite{Singer2006}
The energy is degenerate with respect to the time-reversal operator: only clusters with an even number of magnetic terms (\textit{e.g.}, $x_i$, or $x_i m_i m_j$) will have non-zero effective cluster 
interactions. Conversely, properties that change sign under time-reversal (such as the total magnetic moment) will only have non-zero effective cluster interactions for clusters with an 
odd number of magnetic terms (\textit{e.g.}, $m_i$, or $x_i m_j$).

\section{Results and Discussion}

\subsection{Model Parameterization}

We constructed two chemomagnetic cluster expansions for the (Mn,Fe)Ru$_2$Sn Heusler alloy: one to predict the energy, and one to predict the total magnetic moment.
The chemomagnetic cluster expansions for the energy and magnetic moments were paramaterized using 193 symmetrically-distinct magnetic and chemical configurations enumerated with the CASM\cite{VanderVen2010,Thomas2013,Puchala2013,Puchala2016} software package.
The VASP\cite{Kresse1994,Kresse1996a,Kresse1996} electronic structure code was used to calculate the fully relaxed energies and magnetic moments of the enumerated structures; full details of these calculations can be found in Decolvenaere, \textit{et al}.\cite{Decolvenaere2017a}
The resulting cluster expansions utilized 31 and 18 cluster basis functions respectively. These models had 10-fold cross-validation scores of 3 meV and 0.03 $\mu_B$ per formula unit.

Grand-canonical Monte Carlo (MC) simulations were performed to study both the fully equilibrated and the chemically quenched Mn$_{1-x}$Fe$_x$Ru$_2$Sn Heusler alloy.
In the fully-equilibrated simulations, both chemical and magnetic degrees of freedom were allowed to equilibrate at each temperature.
The chemically-quenched simulations were performed in two steps: a high-temperature annealing stage where both chemical and magnetic degrees of freedom were allowed to equilibrate, and a cooling stage where only magnetic degrees of freedom were equilibrated in the presence of a frozen high temperature chemical microstate.

The Curie and N\'eel temperatures of magnetic Ising models are higher than those of an equivalent Heisenberg model
when using identical interaction coefficients.\cite{Domb1962}
In constructing our phase diagrams, we therefore replaced the scalar spin $m_i$ in our chemomagnetic cluster expansion with a magnetic vector $\bm{m}_i = [a_i, b_i, c_i]$ on the unit sphere, and replaced any products of spins with dot products.
The properties of the Ising and Heisenberg models constructed in this manner converge in the low-temperature 
limit.\cite{Kasteleijn1956}
Further details regarding the fitting of the Hamiltonian and the subsequent MC simulations can be found in the supporting information.

\subsection{Bulk Phase Behavior}

\begin{figure}[h]
        \includegraphics[width=0.45 \textwidth]{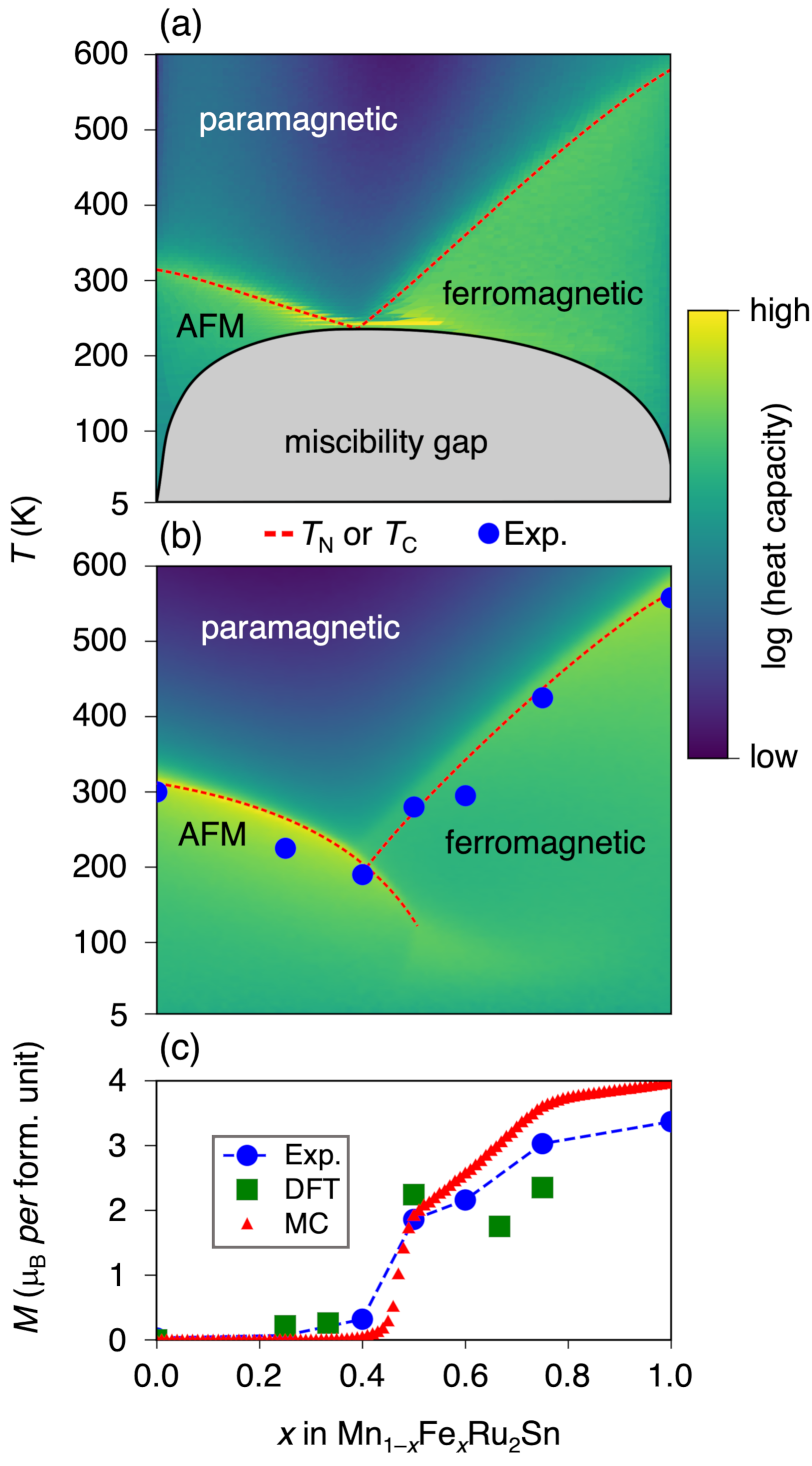}%
        \caption{(a, b) Phase diagrams of (a) the fully-equilibrated and (b)
            chemically-quenched systems. Both plots are shaded by the semi-grand canonical heat capacity ($\frac{\langle \Omega^2 \rangle - \langle \Omega \rangle^2}{k_b T^2}$,  $\Omega = E - (\mu_{Fe} - \mu_{Mn})N_{Fe}$, at a constant magnetic field of 0).
         In both phase diagrams, the red dashed lines indicate approximate magnetic
         (dis)ordering temperatures, based on the heat capacity maxima. In (b),
         the blue dots show the experimental phase
         transitions.\cite{Douglas2016} In the metastable system, for $0.4 < x
         < 0.75$, there is an initial FM ordering followed by local AFM
         ordering upon cooling. (c) Low-temperature net magnetic moment of the
         metastable system at 5\,K and no field, compared to \textit{ab-initio}
         calculations\cite{Decolvenaere2017a} and experiment (saturated moment at
         4\,K).\cite{Douglas2016}
\label{phase}}
    \end{figure}

Figure~\ref{phase}(a) shows the calculated phase diagram for the Mn$_{1-x}$Fe$_x$Ru$_2$Sn Heusler alloy when both chemical and magnetic degrees of freedom are allowed to equilibrate.
The phase diagram shows a paramagnetic (PM) to-FM transition at Fe-rich compositions and a PM-to-AFM transition at Mn-rich compositions, with the magnetic moments of the Mn-rich FCC sublattice adopting the L1$_{1}$ ordering in the AFM phase.
A chemical miscibility gap appears at lower temperatures.
As is evident in Fig.~\ref{phase}(a), the two magnetic transitions and the top of the miscibility gap appear to converge to a chemical-magnetic tricritical point.
A more in-depth analysis of this behavior, however, is beyond the scope of this study.

Figure~\ref{phase}(b) shows the phase diagram when only magnetic degrees of freedom are allowed to equilibrate in the presence of quenched in Mn/Fe disorder.
This phase diagram emulates the conditions of the experimental studies of Douglas \textit{et al}.\cite{Douglas2016}
Figure~\ref{phase}(b) shows that the magnetic transitions are still present in the chemically-quenched Heusler alloy and that they are similar to those appearing in the fully equilibrated phase diagram of Fig.~\ref{phase}(a) for $x < 0.4$ and $x > 0.75$.
The chemical miscibility gap, however, is absent and important differences emerge between the quenched and fully equilibrated Heuslers for $0.4 \leq x \leq 0.75$.
It is in this composition interval that the local magnetic orderings transition smoothly from AFM to FM.
This is shown in Fig~\ref{phase}(c), which compares the calculated magnetization at 5\,K with that measured by Douglas \textit{et al}.\cite{Douglas2016} at 4\,K.
The agreement between calculated and measured magnetizations is very good, although the experimental magnetization above $x \geq 0.75$ is slightly lower than the calculated one, due to the presence of multiple grains and imperfectly aligned magnetic domains in the experimental samples.
The strong overall agreement in Fig~\ref{phase}(c) indicates that the magnetic and chemical microstates sampled with the MC simulations are representative of those in the experimental samples.

    \begin{figure}
        \centering
        \includegraphics[width=0.45 \textwidth]{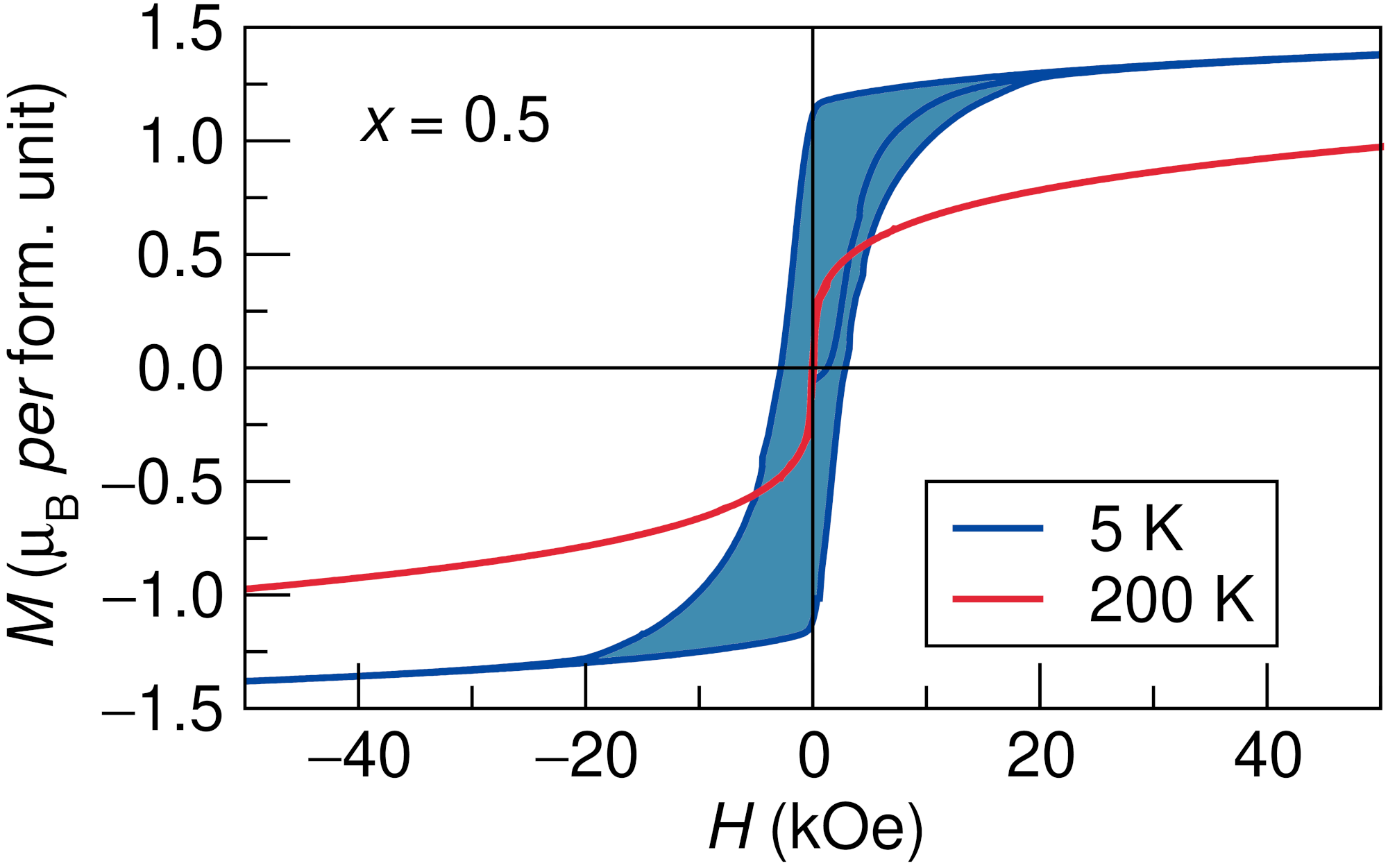}%
        \caption{Experimental magnetization, $M$, as a function of magnetic field, $H$, for $x=0.5$ at 5\,K and 200\,K. 
Details of sample preparation and measurement can be found in reference \cite{Douglas2016}.\label{hyst}}
    \end{figure}

While there is no clear evidence for sharp thermodynamic transitions at intermediate concentrations in the chemically quenched alloy, Fig.~\ref{phase}(b) nevertheless suggests the existence of two qualitatively distinct short-range ordered magnetic states at low temperatures for $0.4 \leq x \leq 0.75$.
This is the composition range where the experimental coercivity increases significantly at low 
temperature\cite{Douglas2016}, even though the solid exists as a disordered solid solution.
Figure~\ref{hyst} compares experimentally measured magnetization curves of Mn$_{1-x}$Fe$_x$Ru$_2$Sn at $x=0.5$ and clearly shows the emergence of magnetic coercivity upon cooling from 200\,K to 5\,K.
Since the chemical degrees of freedom are quenched, this remarkable change in behavior must arise from qualitative changes in the magnetic state with decreasing temperature, as suggested by the calculated phase diagram of Fig.~\ref{phase}(b).

\subsection{Micro-scale Phase Behavior}

The unique occurrence of exchange hardening in a disordered single phase, as shown in Fig.~\ref{hyst}, and the pecular composition dependence of the magnetization of Fig~\ref{phase}(c) has its origin in the strong coupling between magnetic degrees of freedom and local composition.
The peculiarities emerge when there are local fluctuations in the composition relative to the average composition of the solid.
This can be understood by analyzing the behavior of short-range order parameters defined as
    \begin{equation}
        \eta_x = \frac{1}{6} \sum_{j \in \text{NNN}} \frac{x_j + 1}{2}, \quad \quad
        \eta_m = \frac{1}{6} \sum_{j \in \text{NNN}} m_i m_j . \label{eta}
    \end{equation}
The first order-parameter, $\eta_x \in [0, 1]$, measures the local Fe composition around site $i$, while $\eta_m \in [-1, 1]$ is a measure of local magnetic order.
We choose to track compositions and magnetic moments in the next nearest neighbor (NNN) shell of each site as those local correlations are able to uniquely distinguish PM, FM and AFM ordering: perfect L1$_{1}$ AFM ordering yields $\eta_m = -1$ while perfect FM ordering yields $\eta_m = 1$ (there may exist more complex orderings such that $\eta_m = -1$, but we have not observed them in our simulations). The NNN pair interaction also plays an important role in determining magnetic ordering preferences in Mn$_{1-x}$Fe$_x$Ru$_2$Sn as it captures the Sn mediated superexchange interaction between pairs of NNN Mn that are responsible for AFM ordering at Mn rich concentrations \cite{Decolvenaere2017a}
For the case of nearest-neighbor (NN) interactions, L1$_{1}$ and total magnetic disorder share the same value $\eta^{NN}_m = 0$. Analysis using  $\eta^{NN}_m$ produces similar results to results using $\eta^{NNN}_m$, except diminished in magnitude. 

\begin{figure}[h]
        \includegraphics[width=0.45 \textwidth]{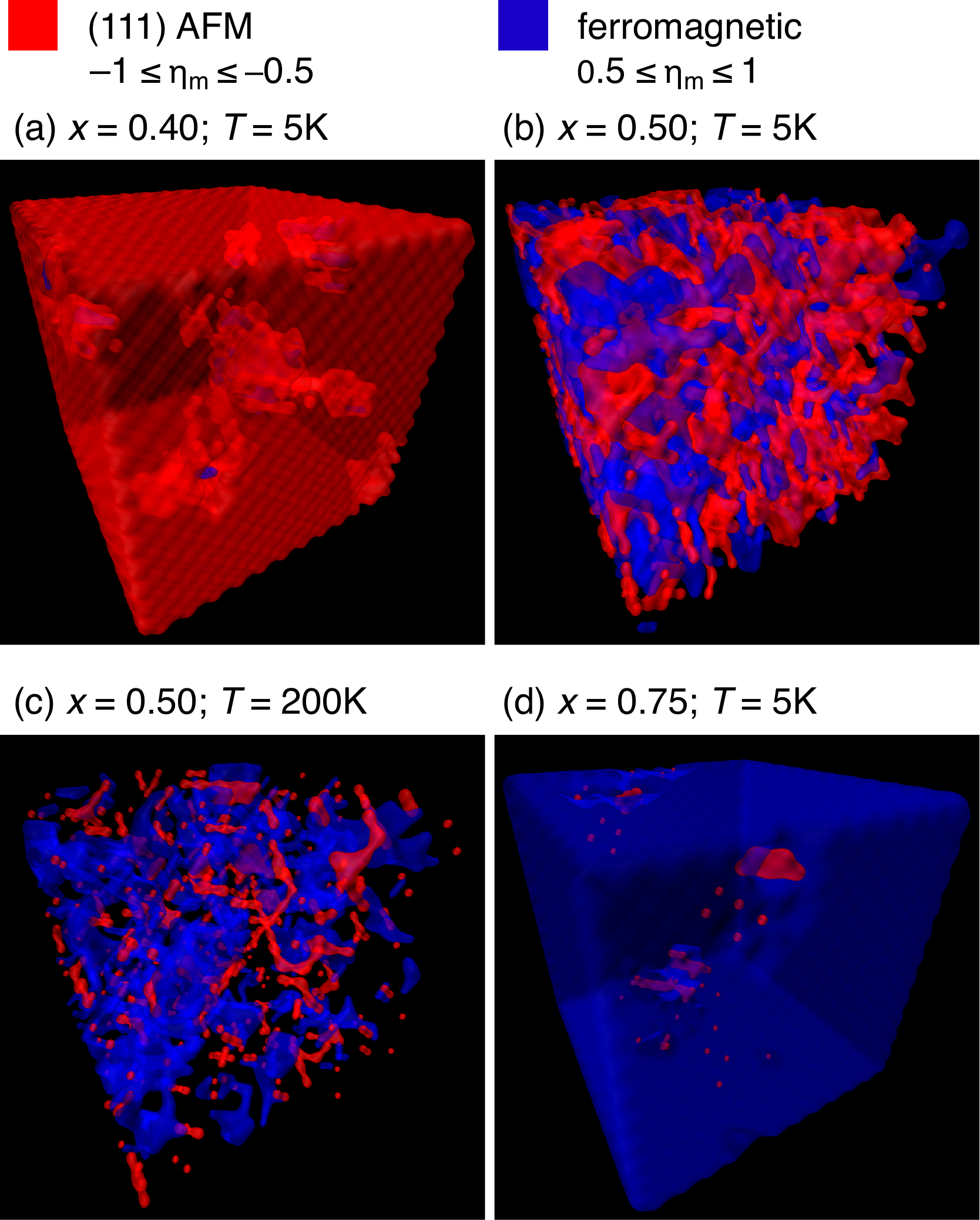}%
        \caption{(a, b, c, d) Snapshots of short-range magnetic orderings in chemically-disordered Mn$_{1-x}$Fe$_{x}$Ru$_2$Sn Heuslers colored by local magnetic order parameter at 5\,K. (c) Snapshot taken at 200\,K of the same simulation of (b). All snapshots represent a 12$\times$12$\times$12 tiling of FCC conventional cells. The white circle indicates a single Mn atom that displaying AFM odering with respect to its neighbors. Additional snapshots can be found in the supporing information.\label{micro}}
\end{figure}

Figure~\ref{micro} shows several MC snap shots that have been color coded to reveal local AFM (red) and FM (blue) regions as measured by the short-range order parameters.
Figure~\ref{micro}(a), (b), and (d), taken at 5 K, show a progressive increase in the phase fraction of FM domains with increasing Fe concentration $x$, consistent with Fig~\ref{phase}(c).
For $x \leq 0.4$, the solid is predominantly AFM with an occasional isolated FM cluster consisting of only a few atomic sites, as illustrated in Fig~\ref{micro}(a).
This changes above $x \geq 0.4$.
At $x=0.5$ for example, (Fig~\ref{micro}(b)) the AFM background becomes permeated by a percolating network of FM regions, all on the scale of only a few atomic sites.
The magnetic moments of the interconnected FM domains are all aligned in the same direction, producing a small, net magnetic moment.
At higher concentrations, the AFM fraction diminishes relative to the FM domains, becoming isolated atomic-scale domains at $x=0.75$ as shown in (Fig~\ref{micro}(d)).

A comparison of Fig~\ref{micro}(b) and (c) shows the effect of temperature on the distribution of AFM and FM domains at $x=0.5$.
While at 5\,K, the FM and AFM regions are compactly intertwined, with sharp transitions in order parameter when going from an FM to AFM region, intervening paramagnetic regions (not color coded) emerge at 200\,K, which form between the FM and AFM domains due to the decreasing thickness of the AFM domains with increasing temperature.
Sharp transitions between FM and AFM regions are known to cause exchange hardening.
Their presence at 5\,K and their absence at 200\,K is consistent with the experimental measurements of Fig~\ref{hyst}.

\begin{figure}[t]
        \includegraphics[width=0.45 \textwidth]{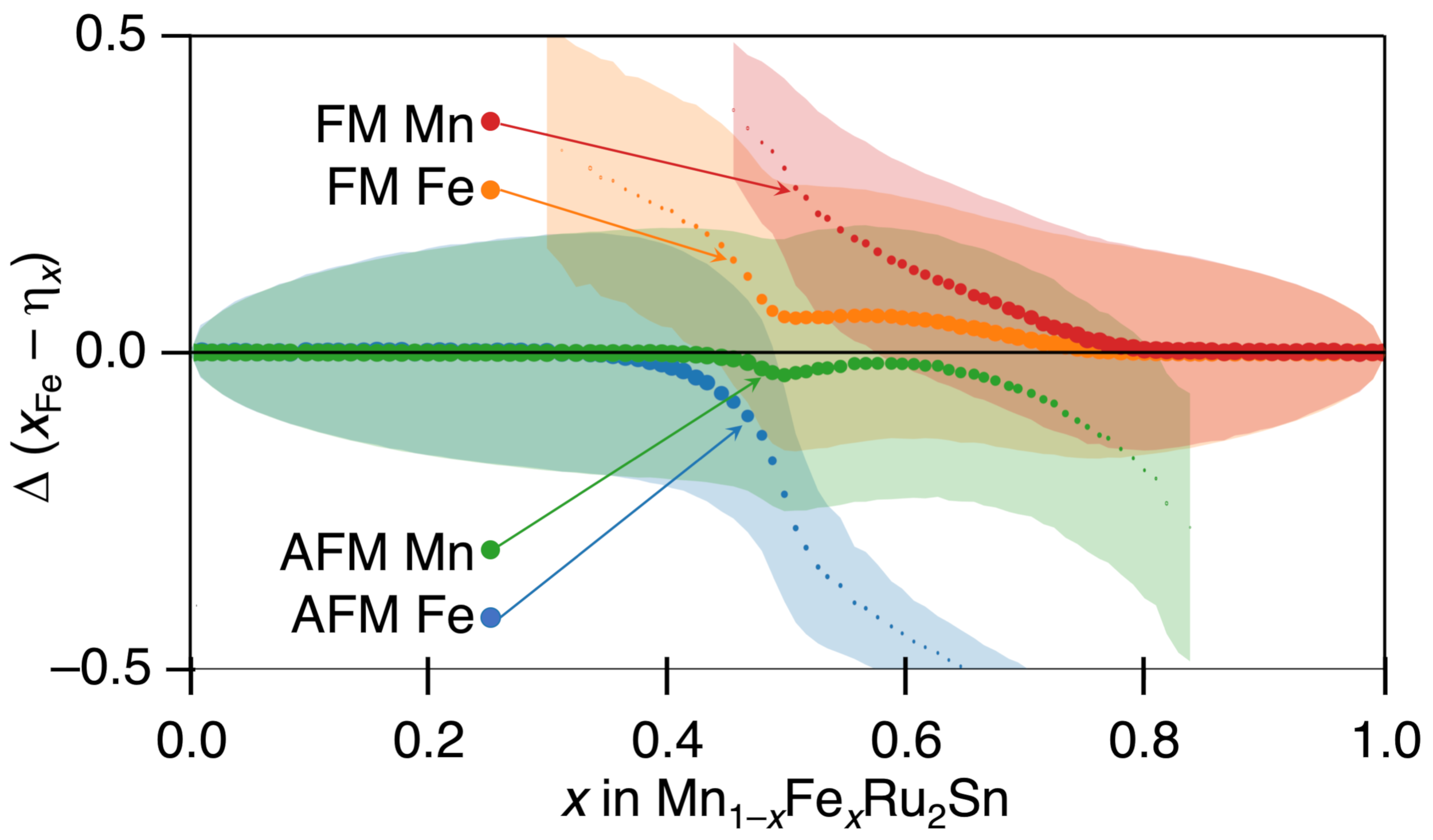}%
        \caption{MC sites at 5 K binned by chemical occupancy (Mn/Fe) and local magnetic ordering (AFM/FM as determined by $\eta_m$). The dots plot the ensemble mean composition deviation, $\Delta$, of all sites belonging to that population (\textit{e.g.}, ``FM Fe''), while the shaded region above and below a dot show the ensemble standard deviation of $\Delta$. The dot size scales with the fraction of total Mn or Fe sites that dot represents with a cutoff of 0.01; not all populations exist at all compositions, \textit{e.g.}, FM Mn for $x \leq 0.5$.\label{dev}}
    \end{figure}

The spatial variations of magnetic ordering, as revealed in the MC snap shots of Fig~\ref{micro}, are strongly correlated to the local concentration.
In fact, the alignment of a magnetic moment at a particular site depends not only on the identity of the occupant of that site (Mn or Fe), but also on the local composition of surrounding sites.
Local fluctuations in concentration that deviate from the average alloy concentration play a particularly important role in determining the unique atomic scale magnetic structures of Fig~\ref{micro}.
This becomes evident upon inspection of Fig.~\ref{dev}, which conveys how local magnetic ordering surrounding Fe and Mn correlates with the local and global composition.
The system is broken up into Mn and Fe-occupied sites, which are further divided into populations of AFM and FM species depending on whether $\eta_m \leq -0.5$ or $\eta_m \geq 0.5$, respectively.
The four populations are plotted in Fig.~\ref{dev}, with the vertical axis measuring the deviation of the local concentration from the average according to $\Delta = x - \eta_x$, while the horizontal axis tracks the average concentration $x$.
The size of each point in Fig.~\ref{dev} reflects the concentration of each population.

Figure~\ref{dev} reveals significant qualitative differences in behavior between Mn and Fe-occupied sites.
Fe sites are strictly AFM until $x=0.35$, then transition rapidly across a narrow composition range to being purely FM at $x=0.6$.
Mn sites, in contrast, exhibit mixed magnetic behavior across a wider range of compositions, with FM and AFM populations both existing between $0.4 \leq x \leq 0.8$.
Additionally, Mn sites exhibit a more even distribution between FM and AFM ordering than Fe sites at intermediate composition, as is evident from the ratio of dot sizes.
Figure~\ref{dev} shows that the existence of both FM and AFM Mn and Fe are strongly correlated with fluctuations in local compositions (i.e. $\Delta \ne 0$) between $x=0.35$ and $x=0.75$.

    Because of quenched in chemical disorder, there is no thermodynamic phase transition with respect to the magnetic degrees of freedom as a function of composition.\cite{Brando2016}
Instead, finite-sized pockets of either FM-in-AFM or AFM-in-FM form past a critical composition of $x=0.40$ or $x=0.75$. These pockets form in areas where statistical fluctuations lead to deviations in the local composition from the average composition, i.e. $\eta_x \neq x$.
The region from $0.35 \leq x \leq 0.75$ can be viewed as similar to a Griffiths phase\cite{Griffiths1969, Vojta2010} as the short-range re-ordering of spins in this composition range is strongly affected by quenched disorder of the chemical degrees of freedom.
Sites are inclined towards AFM or FM ordering, similar to the behavior of competing magnetic orderings 
described by Burgy, \textit{et al}.\cite{Burgy2001}

\section{Conclusion}

In conclusion, we have shown how conditions conducive to exchange hardening can arise in a chemically disordered solid solution.
Our analysis links experiments and theory, demonstrating how simple synthesis techniques can generate new magnetic materials that exhibit exchange hardening \textit{without} requiring two-phase coexistence.
We show that quenched-in local fluctuations in composition can produce a spectrum of FM-in-AFM and AFM-in-FM local orderings, akin to a Griffiths phase.
These effects arise without the need for any long-range order on the (Mn,Fe) sublattice, and reproduce experimentally-observed phenomena previously unexplained.
Due to quenched chemical disorder, the AFM or FM regions in the Griffiths-like region \textit{cannot} coalesce into domains that are large enough to be resolved with commonly used characterization methods.
Instead, the FM and AFM orderings exist in fixed locations dependent on the chemical identity of the site, the local composition around the site, and the bulk composition.

\begin{acknowledgments}
The research reported here was supported by the National Science Foundation (NSF) through the 
Materials Research Science and Engineering Center (MRSEC) at UC Santa Barbara DMR-1720256 (IRG-1).
We acknowledge the use of the resources of the Center for Scientific Computing at UC Santa Barbara supported 
by the NSF MRSEC (DMR-1720256) and NSF CNS-1725797, and of the National Energy Research Scientific Computing 
Center (NERSC), a U.S. Department of Energy Office of Science User Facility operated under 
Contract No. DE-AC02-05CH11231.
\end{acknowledgments}

\bibliography{library.bib}

\end{document}